# Decision-making in Livestock Biosecurity Practices amidst Environmental and Social Uncertainty: Evidence from an Experimental Game

Scott C. Merrill[1*], Christopher J. Koliba[2], Susan M. Moegenburg[1], Asim Zia[2], Jason Parker[3], Timothy Sellnow[4], Serge Wiltshire[5], Gabriela Bucini[1], Caitlin Danehy[6], and Julia M. Smith[6]

[1] Department of Plant and Soil Science, University of Vermont, Burlington, VT 05405-0082

[2] Department of Community Development and Applied Economics, University of Vermont, Burlington, VT 05405

[3] School of Environment and Natural Resources, The Ohio State University at Mansfield, Mansfield, Ohio 44691

[4] Nicholson School of Communication, University of Central Florida, Orlando, FL 32816-1344

[5] Department of Food Systems. University of Vermont, 105 Carrigan Drive, Burlington, VT 05405.

[6] Department of Animal and Veterinary Sciences, University of Vermont, Burlington, VT 05405

* Corresponding Author

E-mail: Scott.C.Merrill@UVM.edu (SCM)

# Abstract

Livestock industries are vulnerable to disease threats, which can cost billions of dollars and have substantial negative social ramifications. Losses are mitigated through increased use of disease-related biosecurity practices, making increased biosecurity an industry goal. Currently, there is no industry-wide standard for sharing information about disease incidence or on-site biosecurity strategies, resulting in uncertainty regarding disease prevalence and biosecurity strategies employed by industry stakeholders. Using an experimental simulation game, we examined human participant's willingness to invest in biosecurity when confronted with disease outbreak scenarios. We varied the scenarios by changing the information provided about 1) disease incidence and 2) biosecurity strategy or response by production facilities to the threat of disease. Here we show that willingness to invest in biosecurity increases with increased information about disease incidence, but decreases with increased information about biosecurity practices used by nearby facilities. Thus, the type or context of the uncertainty confronting the decision maker may be a major factor influencing behavior. Our findings suggest that policies and practices that encourage greater sharing of disease incidence information should have the greatest benefit for protecting herd health.

# Key words

Decision Making; Uncertainty Aversion; Risk; Environmental Uncertainty; Social Uncertainty; Biosecurity; Serious Games, Experimental Games; Livestock Disease

# Introduction

The U.S. livestock industry is vulnerable to disease threats (Homeland Security Presidential Directive HSPD-9), with the potential costs of a major epidemic estimated to be in the 10s to 100s of billions of dollars [1, 2]. Detection of a disease reportable to U.S. animal health authorities, such as Foot and Mouth Disease, could trigger international trade regulations and sanctions resulting in the immediate loss of international market demand for the affected sectors [1]. For example, the closure of swine export markets could result in losses of approximately $5B/yr [3, 4]. The effective mitigation of disease threats, through the right balance of information sharing and biosecurity protocol adoption, can help to reduce risk to individual producers, their production chains and industry wide.

Animal livestock producers operate in a complex and uncertain social and environmental landscape in which information about disease and biosecurity is incomplete or kept confidential within small networks. Despite this lack of information, facility managers must make regular decisions regarding farm operations to attempt to keep animals healthy. While disease spread has been examined, for example through the use of epidemiological models [5, 6], the influence of human behavior on animal health and the spread of disease remains under-examined. Few attempts have been made to quantify the complex factors that influence decisions by industry stakeholders, and moreover, how those decisions will impact the spread of disease [but see 7, 8-11].

Biosecurity practice adoption can be expensive and include infrastructure development and upkeep, personnel training, and opportunity costs [12]. These costs are weighed against the perceived risk of contracting disease, and thus, adoption of biosecurity practices must be weighed against the economic return on investment. Yet, determining the bounds on economic return is challenging because biosecurity efficacy differs by disease, is dependent upon decisions of others, and importantly, is only necessary when confronted with disease. Moreover, non-economic factors influence stakeholder decisions to invest in biosecurity [13].

Understanding the rationale behind biosecurity investment decisions requires an examination of economic ramifications as well as social dynamics, and the risk and uncertainty associated with the decisions. Risk perception in the agricultural sciences carries multiple meanings, such as 1) the probability of a known negative outcome (e.g., risk of loss) [14, 15], and 2) uncertainty of the occurrence of a negative outcome [16, 17]. Uncertainty aversion, the preference for selection of something with a known probability over an unknown probability, is behaviorally ubiquitous with rare exceptions [17]. Moreover, evidence suggests that the majority of individuals will pay a premium to reduce uncertainty [17]. Thus, as it pertains to investment in biosecurity, uncertainty aversion theory would suggest increased investment in biosecurity with increased uncertainty.

One method of compartmentalizing uncertainty is by conceptualizing it as either environmental or social uncertainty (also referred to as strategic uncertainty) [18]. Social uncertainty can be defined as uncertainty related to human decisions, actions or strategies, such as uncertainty if an employee will come to work when they are sick. Environmental uncertainty is uncertainty that is not directly attributable to human decisions, actions or strategies, such as weather uncertainty or uncertainty associated with disease virulence. Messick, Allison, and Samuelson [18] suggest the fundamental distinction between response to environmental uncertainty and social uncertainty is that under environmental uncertainty, people are attempting to optimize their utility, i.e., when individuals make decisions in an environmentally uncertain arena, they simply weigh the pluses and minuses. In contrast, when confronted with social uncertainty, individuals are attempting to strategize based on the anticipated or realized behavior of others. Social uncertainty depends upon the beliefs each participant has about the strategies employed by other people or groups [19]. Loewenstein, Bazerman, and Thompson [20] provide evidence that the decision-making process under social uncertainty contrasts sharply with environmental uncertainty largely because of the potential for social dispute. Others suggest that social uncertainty differs because it can be constrained by our experiences with human behavior and social norms. Overall, evidence suggests that social uncertainty is preferred over environmental uncertainty, implying that environmental uncertainty aversion is greater than to social uncertainty aversion [21, 22]. Typically, people will pay more to reduce environmental uncertainty than to reduce social uncertainty, perhaps because we can use previous experiences to better understand and conceptually reduce uncertainty. For example, we don't know that our doctor is using best practices, but our experiences suggest that they are likely behaving well. For purposes of this study, a social entity's response to the possibility of disease incursion (e.g., biosecurity practices) is labeled social uncertainty, whereas disease factors (e.g., contagion rate, disease hosts) are binned as environmental uncertainty. We hypothesize that higher levels of environmental uncertainty and social uncertainty will be positively related to biosecurity investment. Specifically, as environmental and social uncertainty increase, we predict that participants will be willing to invest more in biosecurity practices to reduce the risk of infection, and thus pay a higher premium to reduce the risk of infection. Because social uncertainty is expected to generate less uncertainty aversion than environmental uncertainty, we hypothesize that increased uncertainty in disease incidence domain (e.g. environmental uncertainty) will have a larger effect on biosecurity investment than an equivalent increase in uncertainty in the biosecurity practice domain (social uncertainty).

It has been well established that risk and uncertainty influence adoption of novel agricultural practices [23]. Moreover, Parker et al. [24] demonstrated that uncertainty exists in the animal food disease transmission domain because of the existence of numerous types of diseases, vectors of transmission, contagion, and virulence factors as well as the array of social or strategic responses to the uncertainty surrounding disease transmission.

Currently, neither the U.S. government nor the livestock industry mandate or incentivize sharing of information about endemic disease incidence or prevalence of biosecurity practices among production-level stakeholders. Many stakeholders consider this information to be confidential, or a part of their trade secret. This lack of industry transparency results in uncertainty regarding the presence of disease threats, and thereby challenges the development of effective biosecurity strategies because it is unclear where weaknesses reside in the industry [25]. Broadly, we are interested in exploring the effect of interventions that would increase information sharing among stakeholders, and thus, provide insight to policy makers who seek to weigh the costs associated with investment in risk communication, public information campaigns and "early warning" systems [e.g., 26].

By deepening our understanding of the relationship between perceptions of environmental and social uncertainty and the inherent behavioral factors considered in light of these uncertainties, we may be able to identify and focus interventions, such as risk communication systems, development of information sharing networks or policy interventions, and by doing so, foster improved biosecurity for livestock industries [27, 28].

# Materials and Methods

## Overview

We simulated swine industry dynamics with an eye towards understanding the effect of uncertainty on the willingness to increase investment in biosecurity. The swine industry was selected as the context for this study because the swine industry has experienced recent incursions of virulent diseases such as porcine respiratory and reproductive syndrome and porcine epidemic diarrhea virus [5] and biosecurity efforts across the swine industry sector are inconsistent, providing an opportunity to help guide behavior towards a more resilient system. Additionally, swine production stakeholders understand the need for improved system resilience and, mediated through workshops and focus groups, helped develop and ground truth our serious game design.

Here we used a digital field experiment, or a "serious game" approach to examine behavior amidst environmental and social uncertainty in the swine industry. Serious games (as opposed to recreational games) are increasingly employed as research tools in studies of human behavior and decision-making [29]. Economic experiments, which are analogous to serious games, have been used for decades to better understand uncertainty and decision-making behaviors [16, 30, 31]. We paid participants using a performance-based incentive strategy to increase salience and engagement by the participants [32].

## Participants

Participants were recruited via Craigslist, University listservs, direct emails, posters, and word of mouth. While participants ranged across the public spectrum, a majority were graduate and undergraduate students. All games were played at the University of Vermont, an R1 university in the northeastern U.S. Upon arrival, participants were given a written informational summary of the study and, because they were paid for their participation, asked to sign a payment authorization form.

To minimize pre-game framing effects, all participants viewed the same pre-recorded narrated slideshow, which provided an overview of the hog industry and explained game play and the cash payout. Participants were informed that they would be acting as managers of a hog production facility with a goal to maximize their economic return by balancing two factors: costs associated with implementing biosecurity practices and costs for any losses caused by the existence of the disease on their simulated facility. No specific disease was named, but participants were told that disease transmission was airborne. The University of Vermont Institutional Review Board approved this project and all protocols were followed for an experiment using human participants.

# Experimental Game Design

To approach a reasonable, and yet realistic blend of risk and reward, we created an experiment that manufactured a conflict between the economics of investing in biosecurity and expected loss in the case of facility infections that could be more reflective of a relatively low impact disease, such as porcine respiratory and reproductive syndrome during an outbreak.

The experimental game was written in the R computer language [33] and was played on Microsoft Surface Pro Tablets [34] with touch screen capacity. Game code can be found in Supporting Information Appendix 1 (S1 Appendix). The experimental game simulated a landscape of fifty hog production facilities, each producing 2500 hogs when they are disease-free (Fig 1). Facilities were randomly located on a 17 by 15 gridded landscape, with the participant operating one production facility randomly selected from one of the 30 centermost possible locations on the grid. That is, the participant's facility was centrally located by design. Facility locations were important because distance between facilities was a factor determining disease spread between facilities.

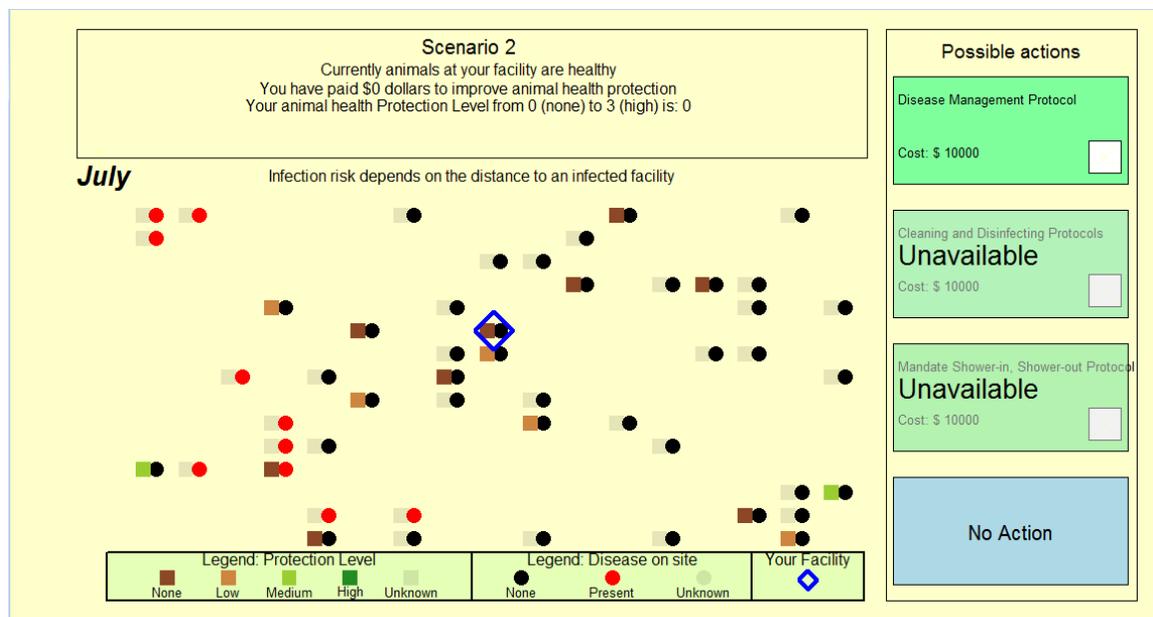

**Fig. 1. Protocol adoption screen capture one.** A screen capture from the Protocol Adoption game showing a map of participant's facility (outlined by blue diamond), 49 other simulation controlled facilities in the region, the scenario (or round) month, each facility's biosecurity adoption level and disease status, and the four possible actions available to the participant.

A blue triangle was used to indicate the participant's facility (Fig 1). The remaining 49 facilities were termed "simulation-controlled facilities." All facilities were shown as circles, the color of which indicated its disease status (black = no disease; red =

disease present; gray = disease status unknown). Next to each circle was a square that indicated the facility's biosecurity level (i.e., none, low, medium, high, or unknown, ranging from dark brown to dark green, respectively).

Participants each played 18 treatment rounds. Each round consisted of 11 biosecurity adoption decisions, simulating monthly decisions from February through December of a single year. In each month, participants made an investment decision regarding biosecurity at their facility. Participants could choose "no action" or they could invest in biosecurity. Only one investment could be made per month. Level 0 was the initial starting point with no biosecurity implemented (0% reduction in the probability of acquiring an infection). Increasing biosecurity to Level 1 by investing in a Disease Management Protocol reduced the probability of infection by 10%. Increasing biosecurity to Level 2 by investing in a Cleaning and Disinfecting Protocol reduced the probability of infection by 40%. Increasing biosecurity to Level 3 by investing in a Shower in/Shower out Protocol reduced the probability of infection by 90%. Even with the maximum biosecurity adopted by the participant (Level 3), their facility's hogs could still become infected, which reflects realistic conditions. Participants were told that a 0 value equated to a Protection Level of "none" and a Protection Level 3 was "high" and were informed that increasing biosecurity by adopting new practices would decrease the probability that their animals would acquire a disease, but they were not informed of the efficacy of each biosecurity practice adoption. Once Level 3 biosecurity was reached, no more choices (except "No Action") were available. Also, if the participant's facility became infected, no more choices (except "No Action") were advised because investing additional funds in biosecurity would not alter their current infection status. Making their choice took participants to either the next month in the current round or, if the month was December, to the next round. Before starting the 18 treatment rounds, participants played two practice rounds. Play during the treatment rounds contributed to the participants' experimental earnings or losses, and generated data that were subsequently analyzed. To reduce the effect of the order of play and any strategic learning that occurred during game play, the 18 treatment scenarios were randomized.

## Social and Environmental Uncertainty Treatments

As an experimental study, the game included two types of treatments designed to represent a change in information sharing that would inform perceptions of uncertainty in two domains: environmental uncertainty and social uncertainty. Each of the two uncertainty treatments had three levels: 1) no information sharing, high uncertainty, 2) partial information sharing, moderate uncertainty, and 3) complete information sharing, low uncertainty. There were three social uncertainty treatments and three environmental uncertainty treatments, comprising nine treatment combinations. Each treatment combination was played twice with biosecurity practices at the simulation-controlled facilities selected by stochastically sampling from each of two different distributions (Table 1). The distribution sets were as follows:
1) Type 1 (High): a distribution with a mean biosecurity level of 2.51. Biosecurity levels for all simulation-controlled facilities were sampled from the set [Level 0, Level 1, Level 2, and Level 3] with respective probabilities of [0.02, 0.05, 0.33, and 0.60].
2) Type 2 (Low): a distribution with a mean biosecurity level of 0.49. Biosecurity levels for all simulation-controlled facilities were obtained by sampling from the set [Level 0, Level 1, Level 2, and Level 3] with respective probabilities of [0.60, 0.33, .05, and 0.02].

**Table 1. Treatment Table.**

|  |  | Degrees of Environmental Uncertainty: Information sharing of disease incidence in the simulation-controlled facilities | | |
|---|---|---|---|---|
|  |  | **Complete** | **Partial** | **None** |
| Degrees of Social Uncertainty: Information sharing of biosecurity levels in the simulation-controlled facilities | **Complete** | High/Low | High/Low | High/Low |
|  | **Partial** | High/Low | High/Low | High/Low |
|  | **None** | High/Low | High/Low | High/Low |

All participants played 18 experimental scenarios: three environmental uncertainty treatments, three social uncertainty treatments, and two distributions of simulation-controlled biosecurity practices (simulation-controlled facilities with relatively high biosecurity practices and simulation-controlled facilities with relatively low biosecurity practices) (Table1).

In the screen shot shown in Fig 2, for example, disease incidence information was not shared with participants, but information about biosecurity practices was completely shared. In this case, biosecurity levels were selected from the Type 1, high biosecurity practice distribution (i.e., most simulation controlled facilities had good to great biosecurity practices in place).

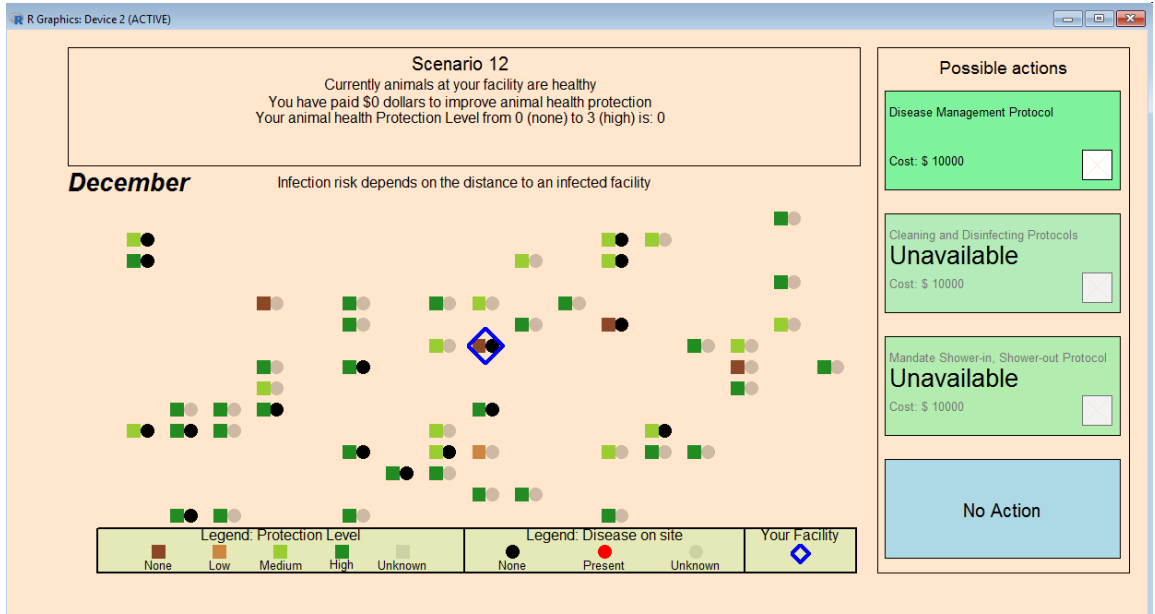

**Fig. 2. Protocol adoption screen capture two.** A screen capture depicting a round with no disease incidence information sharing, complete biosecurity level information sharing, and Type 1 or high biosecurity at simulation-controlled facilities.

## Additional Experimental Game Details

### Disease transmission and probability of infection

Described to the participants as an airborne disease, transmission of disease between all facilities (simulation-controlled and participant-controlled) was a stochastic, probabilistic event with disease spread quantified using a function of distance between facilities and the level of biosecurity protocols adopted by each facility. With a fixed contagion rate, the calculation of the probability of acquiring a new infection at each facility for each time step becomes:

$$1 - \beta_i * \prod_i (1 - \frac{contagion}{distance_{ij}})$$

for each facility i given infected facility j.

Where $\beta_i$ was the biosecurity effect at facility $i$. This calculation provided a single infection probability value for each facility. A random number was then generated and if the random number was lower than the calculated probability value then the facility became infected. We used a *contagion* value of 25. With this contagion value, and no biosecurity adopted by the participant, approximately 75% of rounds with Type 1 biosecurity resulted in an infection at the participant's site, and approximately 15% of rounds with Type 2 biosecurity resulted in an infected site.

At the start of each round, there was a 70% probability of a single infection somewhere among the simulation-controlled facilities. We did not allow the participant's facility to be initialized with an infection, and thus, the participant's facility could not be infected before the participant had had a chance to make a biosecurity practice adoption decision. Additionally, each month there was a 10% chance of a new infection appearing in one of the simulation-controlled facilities.

**Investment economics**

Participants could expect to receive $35,000 experimental dollars minus biosecurity investment costs in gross profit from selling their hogs at the end of the year if hogs were healthy. Monte Carlo simulations were used to generate simulated game play metadata to provide bounds for the range of outcomes that could result from participant decisions. Treatments were each run 80,000 times with stochastic values for infection location and simulation-controlled facility biosecurity levels. When no biosecurity investment was made by the participant, Monte Carlo results indicated that infections would occur at the participant's facility in 31.2% of the scenarios. Gross losses with an infection event were variable but averaged $31,194 experimental dollars. Participants were not told in advance the specific loss they would incur with an infection; rather they were provided an example plus the knowledge that exact losses would depend upon how much of their herd was killed by the disease. Thus, participants were aware that losses would be significant and that they would likely range from around $30,000 to $35,000 experimental dollars. Each incremental increase in biosecurity costs $10,000 experimental dollars, with a maximum investment of $30,000 experimental dollars.

**Relationship between investment decisions, game play and cash payout**

Experimental dollars accrued over the course of the 18 experimental rounds played by each participant. At the end of the game, experimental dollars were converted to real US$ at a rate of $12,000 experimental dollars to $1 US dollar. Participants were paid in cash upon completion of the study. Participants were aware from the outset that their game performance would determine their cash payout, incentivizing them to make decisions that maximized their payout.

**Additional factors that may influence biosecurity adoption behavior**

Behavior in an experimental game situation may be influenced by numerous factors, which were controlled for using the following covariates:

*Temporal discounting:* Psychological distance from an event impacts the perceived value of an event. One example of psychological distance is temporal discounting, which suggests that the value of past or future losses or gains diminishes with increased temporal distance [16, 35, 36]. Because past events are increasingly discounted with temporal distance, we hypothesize that perception of the probability of a future infection event will decrease as time increases since an observation of a past infection event. Specifically, that response effect size, measured by investment in biosecurity, will diminish with time (number of rounds of play) since an infection occurred.

*Order Effect:* While treatment order effect was minimized by randomly selecting the treatment order, participants could change their strategy in a predictable way as the game progressed creating a learning-order effect [37-39]. For example, on average, participants may decrease their adoption of biosecurity practices as the number of rounds increases because the economic return, all other things being equal, favors non-investment strategies.

*Observed Biosecurity Level:* Observed Biosecurity Level at simulation-controlled facilities (categorical: Type 1 high biosecurity, Type 2 low biosecurity, or unknown biosecurity) may affect choice of whether to implement biosecurity.

*Observed Probability of Infection:* We assume some correlation between actual risk of infection based on information available to the participants and their game play. Specifically, as the observable risk increases (e.g., facilities close to the participant's facility become infected), participants should increase their investment in biosecurity. We recognize that substantial variation exists between perceived risk and actual risk, and that perceived risk can be discordant with actual risk. Yet, we anticipated that increases in the actual probability of infection at their facility will correlate with increased investment in biosecurity by the participant.

## Dependent variable calculation

An index variable derived from game data was created that integrates timing and biosecurity practice adoption. The index variable was percent of the maximum amount of biosecurity investment by month, or percent maximum biosecurity (PMB). PMB was calculated by summing biosecurity level by month and dividing it by the sum of the maximum biosecurity investment by month. For example, after the third decision month, if a participant's decisions indicated biosecurity levels of 0, 1, and 2, this would sum to 3. The maximum biosecurity investment possible would be 1, 2, 3, for a sum of 6. Thus, the PMB for the third decision month would be 0.50. Because we were interested in the decision-making process, we viewed PMB data as being informative only when it was possible for participants to make relevant decisions. Relevance could cease under two conditions: 1) when the maximum biosecurity level has been reached or 2) when the facility has become infected. PMB for the month with the last possible decision (The Last Decision Month) was used as the dependent variable in our analysis. PMB was used as an index of biosecurity investment effort because higher PMB values are found with a combination of early and high investment in biosecurity. Last Decision Month is triggered by either reaching the maximum biosecurity level, or by becoming infected. If participants have not maximized their biosecurity level at the end of the round, they received a Last Decision Month value of 13. If Observed Probability of Infection was high, participants may have reached their last decision month sooner by either maximizing biosecurity at their facility or becoming infected. Conversely, if Observed Probability of Infection were low, participants may opt not to invest in biosecurity, resulting in a low PMB. Participants are likely to have an earlier (smaller) last decision month in scenarios with higher perceived or actual risk.

## Data Analysis

Mixed effects linear regression models were used to examine effects of uncertainty treatments and noted covariates. Four mixed effects candidate models were developed to examine the dependent variable, PMB [R version 3.2.0, 33]. Model selection was conducted by selecting the model with the lowest Akaike's Information Criterion (AIC) value [40, 41]. Using the AIC-selected best candidate model, p-values for main effects and interactions were calculated by using Likelihood Ratio Tests [33].

### Population distribution analysis. Environmental and social uncertainty treatments

We analyzed data to examine potential differences between behavioral strategies when confronted with environmental and social uncertainty treatments. Using the selected best candidate model, we generated PMB predictions for each participant for each treatment scenario. Predicted values from the model were then binned by treatment scenario. The binned data were examined as input distributions for two-sample Kolmogorov-Smirnov tests [33]. The Kolmogorov-Smirnov tests quantified the probability that each pair of non-parametric distributions (data from treatment scenario bins) were generated from a single unknown distribution [42].

### Individual differences. Cluster analyses.

Our methodology allows us to look beyond distribution shifts in behavior to examine how individuals in the population responded to changes in uncertainty. Treatment effects look for population level changes in the distribution, whereas individual decisions may contrast with the mean population decisions. We used model prediction data for each treatment for each participant to look for individual responses. Because information sharing is not an industry standard, we define moving from high uncertainty to low uncertainty to be an intervention. Individuals that are receptive to interventions by increasing investment in biosecurity were termed *intervention receptive*., Those that did not change their biosecurity investment behavior based on an intervention were termed *intervention neutral*. Finally, those that react negatively to information sharing interventions, and thus reduce their investment in biosecurity were termed *intervention averse*. We used k-means clustering [33] to bin participants into intervention-response groups labeled as *intervention receptive,* defined as those responding with greater than a 10% increase in biosecurity investment effort (PMB), *intervention neutral,* defined as a minimal change in either direction (+/- 10% change in PMB), or *intervention averse*, defined as greater than a 10% reduction in biosecurity when confronted with scenarios that decreased uncertainty. The elbow technique was used to determine a parsimonious number of behavioral clusters [43].

# Results

Participants (n = 110) took part in the study between February 2016 and July 2016. Of these, 54 identified as female, 52 identified as male, and four did not indicate their sex. Participants ranged in age from 18 – 58 (mean = 23.6). Game play, including the introductory framing slideshow, took approximately one hour. The average payout was $18.20, with a high of $40. Based on game play, many of the participants (61 out of the 110) would have earned less than the research study's minimum payout, so they received the guaranteed minimum $15 participation payment.

## Candidate Model Selection: Mixed effects linear regression models

Of the four models examined, Model 3 was the AIC-selected, best candidate model and all inference was based on it (Table 2).

**Table 2. Mixed-effect linear regression candidate models.**

| Model | Random Variable | EUT | SUT | PI | LM | OBL | TD | OE | LM * PI | EUT * SUT | OBL * SUT | AIC |
|---|---|---|---|---|---|---|---|---|---|---|---|---|
| #1 | Participant | X | X | X | X | X | X | X | X | | | -469.75 |
| #2 | Participant | X | X | X | X | X | X | X | X | X | | -467.28 |
| **#3*** | **Participant** | **X** | **X** | **X** | **X** | **X** | **X** | **X** | **X** | | **X** | **-472.33** |
| #4 | Participant | X | X | X | X | X | X | X | X | X | X | -470 |

Fixed effects included the Environmental Uncertainty Treatment (EUT), Social Uncertainty Treatment (SUT), Observed Probability of Infection (PI), Last Decision Month or the last month where a decision was possible (LM), Observed Biosecurity Level (OBL), Scenarios completed since last infection (Temporal Discounting effect: TD), and round number (Order Effect: OE).
*Akaike's Information Criterion-selected best candidate model.

**Table 3. Model 3 coefficient estimates, standard errors, and t-values**

| Fixed Effect | Estimate | Std. Error | t value |
|---|---|---|---|
| Intercept | 1.1405 | 0.0283 | 40.2330 |
| LM | -0.0872 | 0.0028 | -31.2230 |
| PI | -1.5064 | 0.3545 | -4.2490 |
| EUT (Partial) | -0.0230 | 0.0114 | -2.0120 |
| EUT (Complete) | 0.0484 | 0.0129 | 3.7500 |
| SUT (Partial) | -0.0022 | 0.0139 | -0.1590 |
| SUT (Complete) | -0.0490 | 0.0137 | -3.5830 |
| Type 2 OBL | 0.0111 | 0.0158 | 0.7030 |
| OE | 0.0017 | 0.0009 | 1.8710 |
| TD | -0.0026 | 0.0015 | -1.7500 |
| LM*PI | 0.2271 | 0.0384 | 5.9150 |
| SUT (Partial) * Type 2 OBL | -0.0476 | 0.0222 | -2.1410 |

Random effects: Participant Variance: 0.0184, sd = 0.1355, residual Variance: 0.040, sd = 0.201. Fixed effects included the Environmental Uncertainty Treatment (EUT), Social Uncertainty Treatment (SUT), Observed Probability of Infection (PI), Last Decision Month or the last month where a decision was possible (LM), Observed Biosecurity Level (OBL), Scenarios completed since last infection (Temporal Discounting effect: TD), and round number (Order Effect: OE).

## Population distribution analysis. Environmental and social uncertainty treatments

Based on results from Model 3, both the environmental uncertainty ($\chi^2(1)$=35.75, $p < 0.001$) and social uncertainty ($\chi^2(2)$=8.9265, $p < 0.011$) variables affect PMB (Table 3 and Figs 3 & 4).

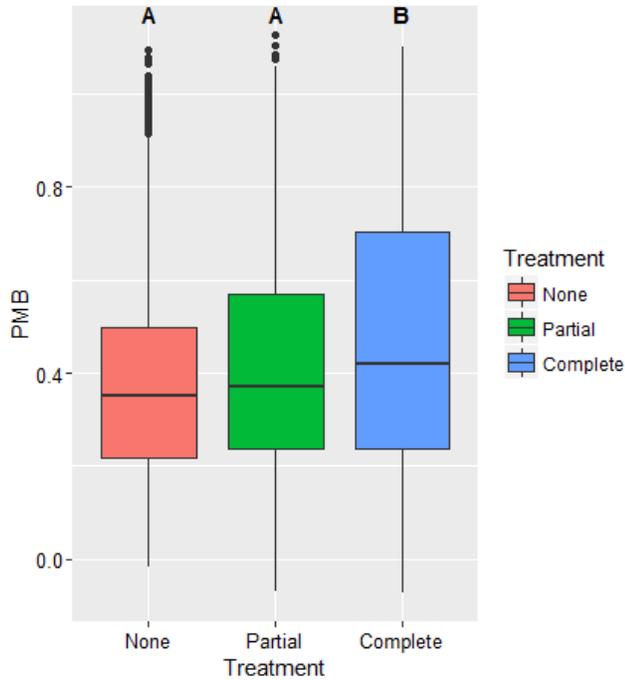

**Fig. 3**. **PMB versus environmental uncertainty.** This figure captures the observed increase in the percent maximum biosecurity (PMB) with increased information sharing and decreased environmental uncertainty associated with disease incidence.

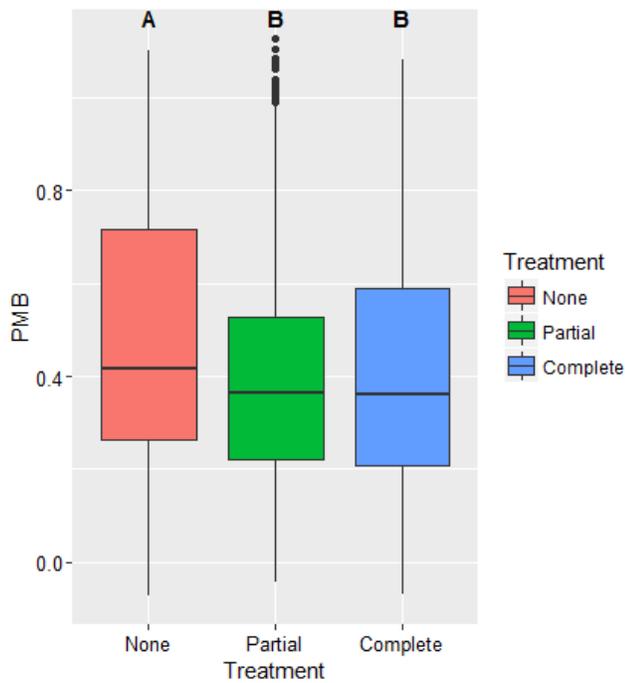

**Fig. 4. PMB versus social uncertainty.** This figure captures the observed decrease in the percent maximum biosecurity (PMB) with increased information

sharing and decreased social uncertainty associated with biosecurity practice reporting.

Table 4 shows differences in the distributions of participant behavior within-treatment using two-sample Kolmogorov-Smirnov tests. The p-values indicate that it is unlikely that the distribution of the PMB values by treatment is derived from the same underlying process.

**Table 4. Results of Kolmogorov-Smirnov tests.**

|  | Environmental Uncertainty (Disease Incidence) | Social Uncertainty (Biosecurity Practice) |
|---|---|---|
| **No I.S. (High Uncertainty) vs Partial I.S. (Moderate Uncertainty)** | $D = 0.061717, p = 0.1621$ | $D = 0.11515, p < 0.001$ |
| **No I.S. (High Uncertainty) vs Complete I.S. (Low Uncertainty)** | $D = 0.1663, p < 0.001$ | $D = 0.12668, p < 0.001$ |
| **Partial I.S. (Moderate Uncertainty) vs Complete I.S. (Low Uncertainty)** | $D = 0.11364, p < 0.001$ | $D = 0.041298, p = 0.627$ |

Tests compare effects of information sharing (I.S.) on participants' decisions by looking for differences in the distributions of participant behavior within the environmental uncertainty treatment and within the social uncertainty treatment. The null hypothesis is that behavioral distributions were drawn from the same unknown distribution [42].

## Individual differences. Cluster analyses.

K-means clustering identified six clusters of participant behavior in both the environmental and social uncertainty treatments using the elbow technique.

Fig 5 describes results from the environmental uncertainty treatments, associated with disease incidence information sharing. Cluster analyses show that 38.2% of participants are *intervention receptive* depicted by those classified into Clusters 5 and 6 with a 0.17 and a 0.34 respective PMB increase. A slightly larger percentage (41.8%) of participants are *intervention neutral*, Clusters 3 and 4, and do not change significantly (-0.06 and 0.06 change in PMB respectively). Interestingly, approximately 20% of individuals, Clusters 1 and 2, are *intervention averse*, showing a decreased willingness to invest in biosecurity, which contrasts with the mean population effect of increased PMB (-0.27 and -0.17 change in PMB).

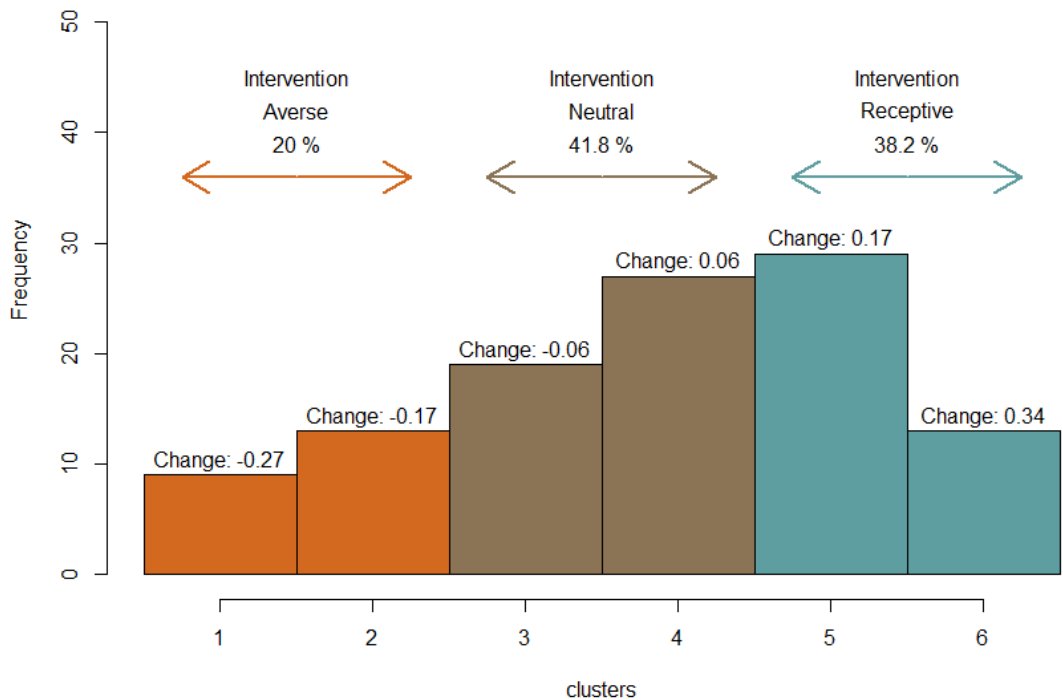

**Fig. 5. Distribution of individual behavior when confronted by changes in environmental uncertainty.** Individual behavior was segregated by observed change from high environmental uncertainty (no disease incidence information sharing) to low environmental uncertainty (complete disease incidence information sharing). While mean population trends suggest increased investment in biosecurity (Fig 3), not all participants followed that relationship. A positive change equates to increased biosecurity investment.

Conversely, cluster analyses results from social uncertainty associated with biosecurity practice information sharing (Fig 6), show that only a small proportion of participants change against the observed mean population trend by increasing their biosecurity investment in response to increased sharing of biosecurity practice information. A small percentage (7.3%) of participants are *intervention receptive* depicted by those classified into a single cluster, Cluster 6 with a 0.19 PMB increase. Fifty percent of participants are *intervention neutral*, Clusters 4 and 5, and do not change significantly (-0.04 and 0.05 change in PMB respectively). A large proportion of individuals, 42.7%, Clusters 1, 2 and 3, are *intervention averse*, changing against the mean obesrved population trend of increased PMB (-0.27 and -0.18 and -0.13 change in PMB).

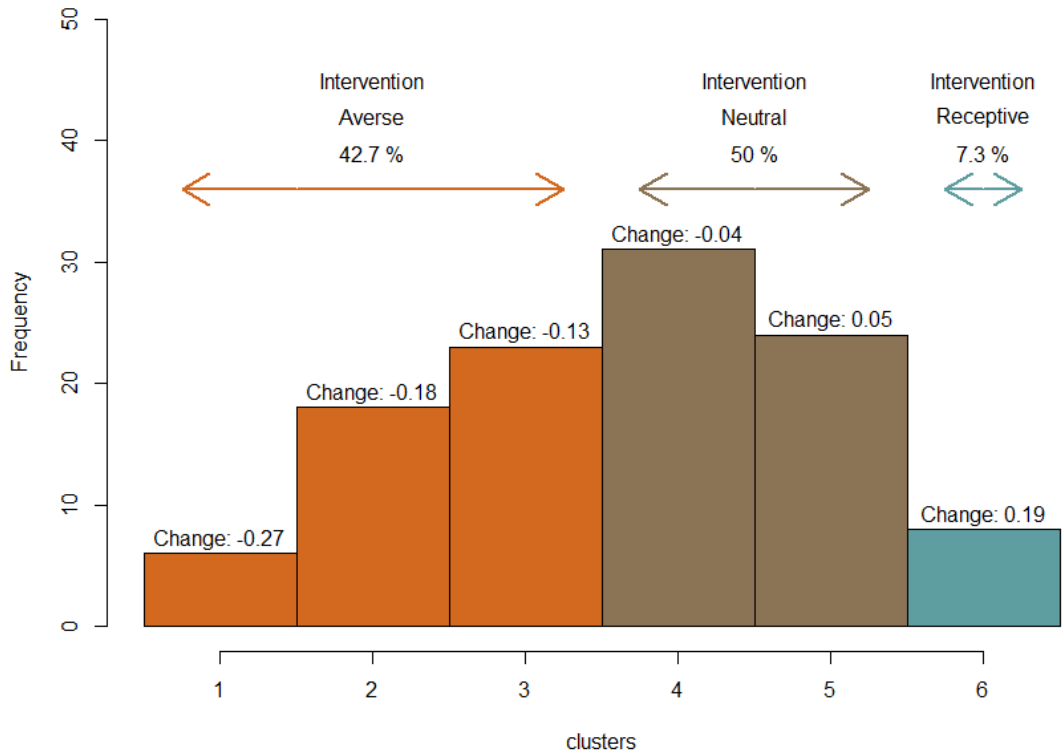

**Fig. 6**. **Distribution of individual behavior when confronted by changes in social uncertainty.** This figure marks individual differences between high social uncertainty (no biosecurity practice information sharing) treatments and low social uncertainty (complete biosecurity practice information sharing) treatments. While mean population trends suggest decreased investment in biosecurity (Fig 4), not all participants followed that relationship. A positive change equates to increased biosecurity investment. There was weak evidence for a temporal discounting effect ($\chi^2(1)=3.012$, $p = 0.083$ and effect size -0.0026). Specifically, experiencing an infection enhances the future perception of risk with participants, reducing the PMB by approximately a quarter of a percentage point every round since they were last infected.

Based on results from Model 3, there was only weak evidence suggesting an order effect with participants increasing their PMB with increased within-game experience (Table 3. $\chi^2(1)=3.483$, $p = 0.062$ and effect size = 0.0017).

Simulation controlled facilities in each scenario were designed to either have relatively high biosecurity (Type 1: mean value observed by participants = 2.53) or relatively low biosecurity (Type 2: mean value observed = 0.46). There is weak evidence ($\chi^2(2)$=5.7009, p = 0.0578) that participants are using the information presented by the level of observed biosecurity reported by simulation-controlled facilities, with more biosecurity investment observed with Type 2 (low) biosecurity than with Type 1 (high) biosecurity (Table 3).

As expected, there is strong evidence that main effects of Last Decision Month and the Observed Probability of Infection and their interaction had a positive effect on PMB ($\chi^2(3)$=1234.8, p < 0.001). The main effects of Last Decision Month (effect size = -0.0872), and Observed Probability of Infection (effect size = -1.5064) and their interaction term (effect size = 0.2271) can be suitably explained by considering of how last decision month is calculated.

## Discussion

We sought to provide insight that will allow policy makers to influence behavior in this social-ecological system towards a more resilient system [44]. Yet, influencing human behavior is complex. One approach is to seek insight into how people might react to disease threats contextualized in the livestock biosecurity arena. Understanding typologies of risk and uncertainty perceptions and decision-making behavior could allow for the "nudging" of human behavior towards resilience. Characteristics of typologies may provide opportunities for better understanding of biosecurity perception including opportunities to increase general awareness and risk acceptance, and improved producer decision-making behavior.

Studies that examine behavioral economics and decision-making frequently select participants from a subset of society, such as a student population. However, these subsets may differ from the population of interest, which carries the potential for bias. And thus, one potential limitation of our findings is that participants in our study were not pre-selected based on knowledge of the industry. Decision-making behavior in the study may differ from stakeholders with extensive experience in the swine industry. For example, U.S. farmers are thought to be less risk averse than the public and much of the discussed risk stemmed from environmental uncertainty [45]. Conversely, Zia et al. [46], using a similar experimental game methodology in the swine industry, did not find a significant difference between swine industry stakeholders and behavior of an audience that was not pre-selected based on experience in the industry. Thus, because the swine production system is complex and decision-makers each have their own set of objectives, any bias that exists may not be consistent.

### Social and environmental uncertainty

Uncertainty aversion theory suggests that in general people will lean towards herd protection, and thus invest more in biosecurity than would be economically optimal in order "to be safe" [17, 31, 47]. Our finding of increased risk mitigation, through investment in biosecurity, with increased social uncertainty generally confirms much of the existing literature [18, 20, 31, 47-49]. While removing social uncertainty through sharing information about biosecurity practices may superficially appear to emerge from participant use of a free-rider strategy, the decrease in biosecurity with reduced social uncertainty exists even when simulation-controlled facilities have very low biosecurity in place (Type 2 biosecurity practice distribution). Thus, simulation-controlled facilities did not provide protection to the wider system, and by inference, the participant, which invalidates the notion that participants were solely relying on a free-rider strategy.

Contrary to the social uncertainty findings, increased environmental uncertainty (e.g. less information about disease prevalence) was correlated with decreased biosecurity investment. Brashers and Hogan [50] argue that in some situations, reduction in uncertainty can produce anxiety, and thus generate a perverse response to uncertainty. Thus, an alternative explanation for our results is that participants were responding with uncertainty preference, optimism or wishful thinking to the environmental uncertainty [17, 22, 49-51]. Gangadharan and Nemes [22] suggest that subjects may underestimate the probability of a negative event if event avoidance is the desired outcome. This would suggest that participants may reduce their investment in biosecurity with increased environmental uncertainty because of some bias that allows for a belief that the negative effect is unlikely to happen to them. One possibility is that perceptions of disease incidence are weighted towards an optimistic belief that facilities were uninfected if no information is provided. That is, the optimistic perception may be that disease is relatively unlikely, and thus, a default hope is that simulation-controlled facilities are likely to be disease-free. Our findings would suggest that participants held onto an optimistic perception that unknown status of others' facilities implied healthy facilities.

While we hypothesized differences between social and environmental uncertainty, previous research did not lead us to expect that parsing of uncertainty into response behavior associated with social uncertainty and environmental uncertainty would elicit diametric biosecurity investment responses. Loewenstein et al. [20] found behavioral distinction between decisions made under social uncertainty compared with environmental uncertainty situations. Their results showed large effect size changes, yet relationships between social and environmental uncertainty treatments remained similar, showing uncertainty aversion throughout their experiments, and thus unlike our results, did not observe uncertainty preference in any of their treatments. We encourage future research to extend the understanding of uncertainty aversion by determining the role of the uncertainty context on decision-making behavior.

Perhaps unsurprisingly, participant's investment in biosecurity increased as the observable risk of infection increased at a participant's facility, such as when a neighboring facility became infected. The timing of this increased risk was also important, with increased risk towards the end of a round less likely to promote investment than the same risk early in the round. This suggests that most individuals were responding appropriately to increased relative risk of infection in their environment because a relatively small risk will compound from early scenarios to late scenarios possibly resulting in a high probability of infection, whereas the same small risk in a late round will result in a relatively small probability of infection by the end of the round.

There was weak evidence for a temporal discounting effect. In general, participants responded to an infection with an initial increase in biosecurity investment followed by a gradual decline. Evidence for a temporal discounting effect is supported by the literature [35, 36, 52] and has been observed in food animal industries, such as the poultry industry, with heightened biosecurity during an outbreak and just subsequent to outbreaks.

Variability associated with the play order effect was controlled using a learning effect variable. Participants became increasingly risk averse as the game progressed, possibly associated with loss aversion as their experimental dollars accrued [53, 54].

Overall, results provide suggested direction for policy makers and the development of new interventions, yet, a conservative approach to altering policy may be warranted. Our participants, while recruited from the public, likely did not have experience in the swine industry, and thus, may differ behaviorally from those working with hogs. Moreover, we created an elevated risk of infection to create a conflict requiring action: Experimental dollars were reflective of reality as far as expected profit per hog, yet did not accurately represent the time requirements for raising a hog, the costs for implementing new biosecurity practices, upkeep of biosecurity practices, or the expected loss if a disease were acquired at a hog production facility. Reality appears to reflect a much lower chance of infection, with a much higher cost when an infection occurs, inclusive of the possibility of loss of livelihood [1], a set of contextual factors that can be studied in future iteration of the experimental game.

## Conclusion

With these results, we gain further understanding that increased communication and sharing of information might not always result in increased biosecurity. Rather, communication strategies must be designed thoughtfully and with careful audience analysis. Insight into uncertainty, risk and the importance of their context should provoke discussions about the ramifications of potential interventions, with the hope that well-considered interventions will generate measurable change in biosecurity attitudes and intent of stakeholders. Our findings challenge the assertion that simply increasing the communication of information regarding biosecurity strategy and implementation in the industry will result in increased investment in biosecurity. Willingness to invest in heightened biosecurity increases with increased awareness of disease incidence in the system, but decreases with increased awareness of biosecurity practices in place at nearby facilities. Thus, recommendations designed to nudge behavior towards a more biosecure industry require an understanding that response to an intervention that decreases uncertainty of disease prevalence may be more effective than decreasing uncertainty of biosecurity strategy adoption. Our results suggest that policy makers aiming to enhance industry biosecurity may be successful by communicating the incidences of diseases and infection outbreaks to livestock owners/managers. In contrast, dissemination of information about biosecurity practices used at other facilities should not be actively encouraged, because these results suggest that information sharing in this domain may result in decreased investment in biosecurity.

# Acknowledgements

This material is based upon work that is supported by the National Institute of Food and Agriculture, U.S. Department of Agriculture, under award number 2015-69004-23273. Any opinions, findings, conclusions, or recommendations expressed in this publication are those of the author(s) and do not necessarily reflect the view of the U.S. Department of Agriculture.